\documentclass[%
 twocolumn,
 amsmath,
 aps, pra,
]{revtex4-1}

\usepackage{color}
\usepackage{graphicx}
           
\renewcommand{\Re}{\mathop{\mathrm{Re}}\nolimits}           
\renewcommand{\Im}{\mathop{\mathrm{Im}}\nolimits}           
            
\begin{document}

\title{Role of disorder in plasmon-assisted near-field heat transfer between two-dimensional metals}

\author{Jonathan L. Wise}
\email{jonathan.wise@lpmmc.cnrs.fr}
\affiliation{Universit\'e Grenoble Alpes and CNRS, LPMMC, 25 rue des Martyrs, 38042 Grenoble, France}

\author{Denis M. Basko}
\affiliation{Universit\'e Grenoble Alpes and CNRS, LPMMC, 25 rue des Martyrs, 38042 Grenoble, France}

\author{Frank W. J. Hekking}
\altaffiliation[]{Deceased 15 May 2017}
\affiliation{Universit\'e Grenoble Alpes and CNRS, LPMMC, 25 rue des Martyrs, 38042 Grenoble, France}

\begin{abstract}
We perform a theoretical study of the near-field heat transfer between two atomically thin metallic layers, isolated galvanically but coupled by the Coulomb interaction, within the framework of fluctuational electrodynamics in the Coulomb limit. 
We clarify the role of disorder and spatial dispersion, and identify several distinct regimes of the heat transfer.
We find that the plasmon contribution to the heat current is suppressed in both the clean and diffusive limits, but dominates in a parametrically wide crossover region at sufficiently high temperatures.
\end{abstract}

\maketitle

\section{Introduction}
\label{sec:intro}

In metals, heat is most efficiently transported by electrons, as manifested by the Wiedemann-Franz law. In the absence of a galvanic contact between two nearby metals, heat can be transferred by electromagnetic fluctuations~\cite{Rytov1953, Polder1971, Levin1980, Loomis1994, Pendry1999}. In the near field the heat flux can be strongly enhanced compared to the Planckian radiative one, as was observed in many experiments~(see reviews \cite{Joulain2005, Volokitin2007, Song2015} as well as Refs.~\cite{Kim2015, Song2016, StGelais2016, Kloppstech2017, Cui2017, Yang2018, Sabbaghi2019} for more recent experiments). Depending on the circumstances, such contactless heat transfer can be useful or harmful.

The standard framework for the theoretical description of this heat transfer is fluctuational electrodynamics~\cite{Polder1971, Rytov1989}. It provides a very intuitive picture: the transferred heat can be viewed as energy dissipated in one part of the structure by fluctuating electromagnetic fields, which are created by fluctuating charges and currents in the other part of the structure. According to the fluctuation-dissipation theorem, both dissipation and fluctuations are determined by the material response functions (such as optical conductivity), which thus serve as the input to this phenomenological theory. The resulting heat current is then entirely determined by the system geometry and the model used for the material response.
For the latter, models with different levels of sophistication have been used in the literature. They range from assuming a local frequency-dependent conductivity~\cite{Polder1971}, to including non-local effects via spatial dispersion~\cite{Chapuis2008} or via surface contributions to the response~\cite{Levin1980, Volokitin2001}. From the literature, it is often difficult to decide which ingredients are important in which situation.

In particular, in several works dedicated to different materials in the near-field regime, the important role played by collective plasmon excitations has been pointed out~\cite{Rousseau2009, Svetovoy2012, Ilic2012, Rodriguez2015, Yu2017, Jiang2017, Zhao2017, Zhang2019, Wang2019}. However, Ref.~\cite{Mahan2017} concluded that surface plasmons were unimportant for the heat transfer between two bulk semi-infinite metals, and in Ref.~\cite{Kamenev2018} no plasmon contribution was reported for heat current between two thin metallic layers in the clean and diffusive limits.

To clarify this issue, we revisit the problem of heat transfer between two thin parallel metallic layers, assuming the separation between them to be
small enough, so that retardation effects can be neglected and the transfer is dominated by the Coulomb interaction (the conditions for this are discussed in Sec.~\ref{ssec:results}).
Including disorder and spatial dispersion within the framework of fluctuational electrodynamics in the Coulomb limit, we identify different regimes of interlayer separation and temperature where different physical ingredients are important. We find that the plasmon contribution to the heat current is suppressed in both the clean and diffusive limits, but dominates in a parametrically wide crossover region at sufficiently high temperatures. 
We also discuss the applicability of the effective circuit approach~\cite{Pascal2011} to heat transfer between metals.

The paper is organised as follows.
In Sec.~\ref{sec:model} we specify our model for the system, discuss the main regimes of heat transfer, present the resulting expressions for the heat current, and discuss their sensitivity to the specific model chosen here.
We give the detailed derivation of the results in Sec.~\ref{sec:results}, and verify them numerically in Sec.~\ref{sec:numerics}. In Sec.~\ref{sec:circuit}, we interpret our results in terms of circuit theory, 
and in Sec.~\ref{sec:experiments} we discuss some relevant experiments.
In Appendix~\ref{app:response}, we give a derivation of the density response function for a two-dimensional (2D) metal with short-range impurities, interpolating between clean and diffusive limits.

\section{Model and summary of results}
\label{sec:model}
\subsection{Model}
\label{ssec:model}
We consider two thin parallel metallic layers, labeled 1~and~2, separated by a distance~$d$, and kept at different temperatures $T_1$ and $T_2$. 
We assume to be in the limit $d\ll{c}/T_{1,2}$, where $c$~is the speed of light, and we use units where Planck and Boltzmann constants are set to 1. In this limit, heat transfer is dominated by the Coulomb interaction.
Our analysis is based on the standard expression for the heat current per unit area~\cite{Yu2017, Jiang2017, Zhang2018, Wang2018, Kamenev2018},
\begin{align}
&J(T_1,T_2) = \int_0^\infty \frac{d\omega}\pi
\int\frac{d^2\mathbf{q}}{(2\pi)^2}\,
\omega\left[\mathcal{N}_1(\omega)-\mathcal{N}_2(\omega)\right]
\mathcal{T}(\mathbf{q},\omega), \label{eqn:heat}\\
&\mathcal{T}(\mathbf{q},\omega)=
2\left|V_{12}(\mathbf{q}, \omega)\right|^2 \Im\Pi_1(\mathbf{q}, \omega)\Im\Pi_2(\mathbf{q}, \omega), \label{eqn:transmission}
\end{align}
often called the Caroli formula, in analogy to a similar expression for electron current across a tunnel junction~\cite{Caroli1971}. In Eq.~(\ref{eqn:heat}), $\mathcal{N}_{1,2}(\omega)=1/[\exp(\omega/T_{1,2})-1]$ are the Bose distribution functions in the two layers. In Eq.~(\ref{eqn:transmission}),
\begin{equation}\label{eqn:V12}
V_{12}(\mathbf{q},\omega)=\frac{v_\mathbf{q}e^{-qd}}{(1-v_\mathbf{q}\Pi_1)(1-v_\mathbf{q}\Pi_2)-v_\mathbf{q}^2\Pi_1\Pi_2e^{-2qd}}
\end{equation}
is the Coulomb interaction between the layers, screened in the random-phase approximation, with $v_\mathbf{q}=2\pi{e}^2/q$ being the bare 2D Coulomb potential and $e<0$ the electron charge. Finally, $\Pi_1(\mathbf{q},\omega)$ and $\Pi_2(\mathbf{q},\omega)$ are the susceptibilities determining the linear response $\delta\rho_i(\mathbf{r},t)=\Pi_i(\mathbf{q},\omega)\,e\varphi_{\mathbf{q},\omega}\,e^{i\mathbf{q}\mathbf{r}-i\omega{t}}$ of the 2D electron density $\rho_i$ in the corresponding layer $i=1,2$ to a perturbing electrostatic potential $\varphi_{\mathbf{q},\omega}\,e^{i\mathbf{q}\mathbf{r}-i\omega{t}}$ applied to that layer. The density response function is related to the in-plane longitudinal optical conductivity, $\sigma(\mathbf{q},\omega)=(i\omega/q^2)e^2\Pi(\mathbf{q},\omega)$.
Equation~(\ref{eqn:heat}) can be derived from the Coulomb limit of fluctuational electrodynamics~\cite{Yu2017}, from non-equilibrium Green's function formalism~\cite{Zhang2018}, or from the kinetic equation~\cite{Kamenev2018}.

The density response functions encode all material characteristics of the two layers. We model each layer as a degenerate isotropic 2D electron gas with short-range impurities. Such a system can be characterised by three parameters: (i)~$\nu$,~the electronic density of states per unit area at the Fermi level including both spin projections, whose energy dependence is neglected; (ii)~$v_F$,~the Fermi velocity; and (iii)~$\tau$, the elastic scattering time. The Fermi momentum $p_F$ is assumed to be the largest momentum scale, $p_F\gg{q},\,\omega/v_F,\,T/v_F,\,1/(v_F\tau)$. Under these assumptions, the density response function of each layer is temperature-independent and given by~\cite{Zala2001}
\begin{equation}\label{eqn:polarizability}
\Pi(\mathbf{q},\omega)
=-\nu\left[1+\frac{i\omega\tau}{\sqrt{(1-i\omega\tau)^2+(v_Fq\tau)^2}-1}\right],
\end{equation}
for an arbitrary relation between $\omega$, $v_Fq$, and~$\tau$. Equation~(\ref{eqn:polarizability}) interpolates between two well-known expressions corresponding to the clean limit ($\tau\to\infty$) and the diffusive limit ($\omega, v_Fq\ll{1}/\tau$):
\begin{subequations}\begin{align}
&\Pi(\mathbf{q},\omega)=-\nu\left[1+\frac{i\omega}{\sqrt{(v_Fq)^2-\omega^2}}\right]
\;\;\;\mbox{(clean)},
\label{eqn:Piclean}\\
&\Pi(\mathbf{q},\omega)=-\frac{\nu{D}q^2}{Dq^2-i\omega}\;\;\;\mbox{(diffusive)}\label{eqn:Pidiff},
\end{align}\end{subequations}
where $D=v_F^2\tau/2$ is the diffusion coefficient.
Equation~(\ref{eqn:polarizability}) corresponds to a 2D analog of Mermin's prescription~\cite{Mermin1970}, which was recently used to model the response in monolayer graphene~\cite{Jablan2009, Svetovoy2012}. In Appendix~\ref{app:response} we give a derivation of Eq.~(\ref{eqn:polarizability}) based on the Boltzmann equation for electrons scattering on impurities. In spite of several assumptions underlying Eq.~(\ref{eqn:polarizability}), its simplicity enables one to obtain an important insight into the interplay between the spatial dispersion [that is, the $\mathbf{q}$~dependence of the conductivity $\sigma(\mathbf{q},\omega)$] and the impurity scattering. For simplicity we also assume the two metals to be identical.

The three independent material parameters, introduced above, define two important length scales: (i)~the 2D screening length $1/\kappa=(2\pi{e}^2\nu)^{-1}$, and (ii)~the mean free path $\ell={v}_F\tau$. Typically, the screening length is very short (on the atomic scale), so we assume $\kappa\ell\gg1$ and $\kappa{d}\gg1$. The mean free path can vary from several nanometers to several microns, and may be larger or smaller than the separation~$d$ between the two layers.
 
\subsection{Summary of results}
\label{ssec:results}
In the isotropic model, formulated above, Eq.~(\ref{eqn:heat}) represents a two-dimensional integral over $q$ and~$\omega$. This integral can be rather straightforwardly evaluated numerically, but much better insight into relevant physics is obtained by studying different asymptotics of the integral in various parameter regimes. The latter approach was adopted in Ref.~\cite{Kamenev2018} using the two limiting expressions (\ref{eqn:Piclean}) and (\ref{eqn:Pidiff}). It turns out, however, that these expressions miss the plasmon contribution.

\begin{figure}
\centering
\includegraphics[width=0.4\textwidth]{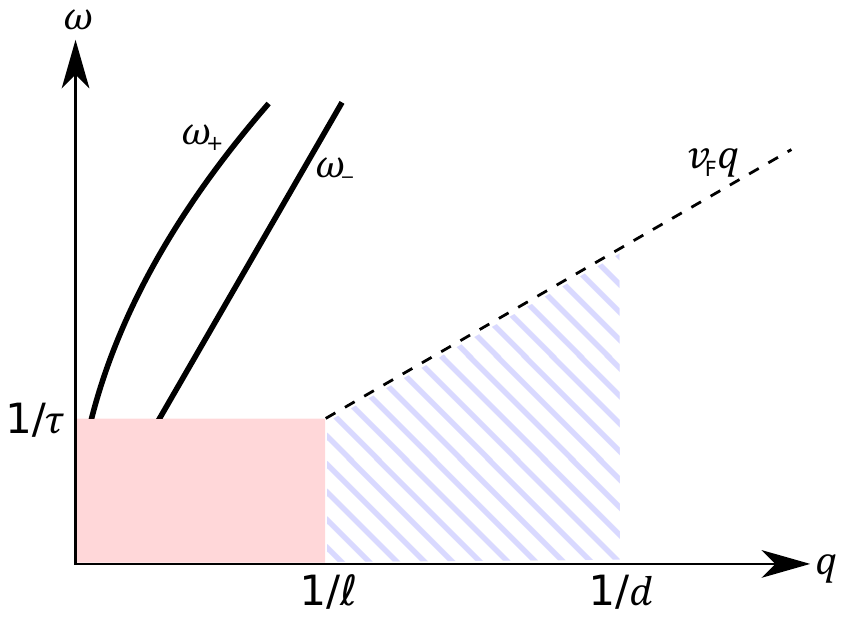}
\caption{A schematic picture showing the behaviour of $\mathcal{T}(q,\omega)$ [Eq.~(\ref{eqn:transmission})] in the $(q,\omega)$ plane for $d\ll\ell$. For $q\gg1/d$ (to the right of the hatched area), $\mathcal{T}(q,\omega)$ is suppressed by the factor $e^{-qd}$. In the hatched area, $\mathcal{T}(q,\omega)$ is well approximated by the clean limit expression~(\ref{eqn:Piclean}), while the shaded area $q\ell\ll1$, $\omega\tau\ll1$ corresponds to the diffusive limit~(\ref{eqn:Pidiff}). In the white region above the shaded and hatched areas ($\omega\gg1/\tau$, $\omega>v_Fq$) the integrand is small except, maybe, in the vicinity of the symmetric and antisymmetric plasmon dispersions (upper and lower solid lines, respectively) where $|V_{12}|^2$ is peaked.}
\label{fig:qw-plane}
\end{figure}

We show schematically the behaviour of of $\mathcal{T}(q,\omega)$ [Eq.~(\ref{eqn:transmission})] in the $(q,\omega)$ plane for $d\ll\ell$ in Fig.~\ref{fig:qw-plane}. At large $\omega\gtrsim\max\{T_1,T_2\}$, the integrand in Eq.~(\ref{eqn:heat}) is suppressed by the Bose function at $\omega$, and this cutoff may be positioned anywhere in Fig.~\ref{fig:qw-plane}, depending on the temperatures. At large~$q\gtrsim1/d$, the integrand is suppressed by $e^{-qd}$ in the numerator of Eq.~(\ref{eqn:V12}); Fig.~\ref{fig:qw-plane} corresponds to the case $d\ll\ell$, but for larger $d$ the spatial cutoff may shift to the diffusive shaded area.

The strongly coupled plasmon modes (in the case of identical layers, symmetric and antisymmetric, denoted by ``$\pm$'') manifest themselves as poles of $V_{12}(q,\omega)$. In the clean limit, $\tau\to\infty$, the plasmon frequencies are real and given by (for $q\ll\kappa$)
\begin{equation}\label{eqn:plasmons}
\omega_\pm=v_F\sqrt{\kappa{q}\,\frac{1\pm{e}^{-qd}}2}.
\end{equation}
At finite $\tau$, but such that $\omega_\pm\tau\gg1$, the poles acquire a small imaginary part, so $|V_{12}(q,\omega)|^2$ is peaked around the dispersions~(\ref{eqn:plasmons}). At $\omega_\pm\tau\lesssim1$, when the diffusive expression~(\ref{eqn:Pidiff}) applies, the plasmons are overdamped and do not produce a separate contribution to the integral. In the strictly clean limit, $\tau\to\infty$, their contribution vanishes as well, since for $\omega>v_Fq$ the integrand vanishes because $\Im\Pi(q,\omega>v_Fq)=0$.

\begin{figure}
\centering
\includegraphics[width=0.48\textwidth]{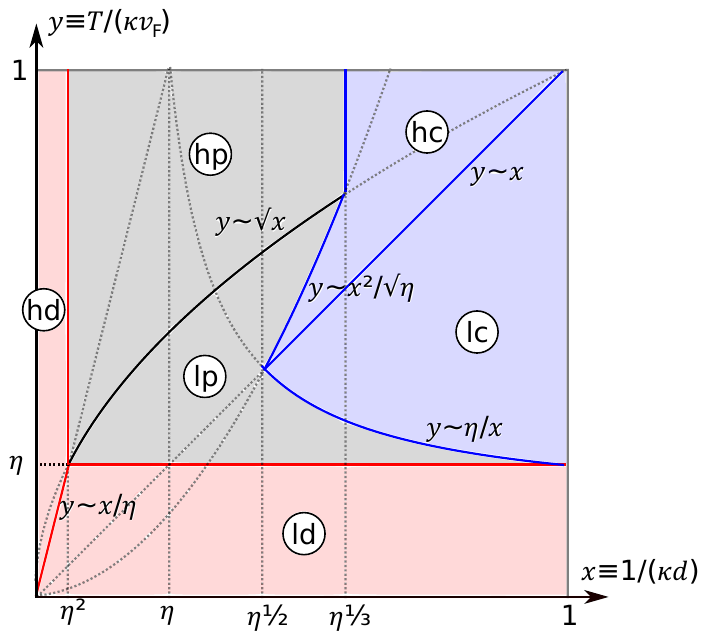}
\caption{The regions in which clean (blue), plasmonic (grey) and diffusive (red), contributions are dominant in the heat current and the domains of validity for asymptotic expressions~(\ref{eqn:results}) in the parameter plane $(1/d,T)$, shown in the dimensionless variables $x\equiv1/(\kappa{d})$, $y\equiv{T}/(\kappa{v}_F)$. The boundaries between the regimes are governed by a single dimensionless material parameter $\eta=1/(\kappa\ell)\ll1$.}
\label{fig:domregions}
\end{figure}

For temperature-independent $\Pi(\mathbf{q},\omega)$, Eq.~(\ref{eqn:heat}) naturally splits into the difference $J(T_1,T_2)=J(T_1)-J(T_2)$.
A detailed analysis of different asymptotic regimes for the integral in Eq.~(\ref{eqn:heat}) (given in the next section) results in several asymptotic expressions for $J(T)$:
\begin{subequations}\label{eqn:results}\begin{eqnarray}
&& J_{\rm lc}(T)=\frac{\pi^2}{60}\, \frac{T^4}{v_F^2(\kappa d)^2}\,
\ln\frac{v_F}{Td},\label{eqn:Jlc}\\ 
&& J_{\rm hc}(T)=\frac{\pi^2}{900}\, \frac{v_F}{d^3}\frac{T}{\left(\kappa d\right)^2},\label{eqn:Jhc}\\ 
&& J_{\rm lp}(T)=\frac{\zeta(3)}{4\pi}\,\frac{T^3}{D\kappa d},
\label{eqn:Jlp}\\ 
&& J_{\rm hp}(T)=\frac{T}{16\pi\tau{d}^2}\,
\mathcal{L}(\ell^2\kappa/4d),\label{eqn:Jhp}\\ 
&& J_{\rm ld}(T)=\frac{\zeta(3)}{8\pi}\, \frac{T^3}{D\kappa d},\label{eqn:Jld}\\  
&& J_{\rm hd}(T)=\frac{1}{16\pi}\,\frac{D\kappa}{d^3}\,T.\label{eqn:Jhd}
\end{eqnarray}\end{subequations}
In the labels ``l" and ``h'' denote low and high temperature, while ``c", ``p", and ``d'' stand for clean, plasmonic, and diffusive, respectively. Here $\zeta(x)$ is the Riemann zeta function, and $\mathcal{L}(x)$ is a slow logarithmic function defined in Eq.~(\ref{eqn:Ldef}). For moderate values of $\ln(\ell^2\kappa/4d)<10$, it can be approximated with 10\% precision as
\begin{equation}\label{eqn:Lapprox}
\mathcal{L}(x)\approx\frac{4\ln^3x}{1+(\ln{x})/\ln(1+\ln{x})}.
\end{equation}
Figure~\ref{fig:domregions} schematically shows the domains of validity for  expressions (\ref{eqn:results}) in the $(1/d,T)$ plane.
The clean and diffusive regimes were also identified in Ref.~\cite{Kamenev2018}.

Each of the above regimes is characterised by typical scales of $q$ and $\omega$, which dominate the integral in Eq.~(\ref{eqn:heat}). Namely, $q\sim1/d$ in the all high temperature cases, $T/(v_F\sqrt{\kappa d})$ in the low temperature plasmonic case and $\sqrt{T/D\kappa d}$ in the low temperature diffusive case.  In the low temperature clean case, the momentum integral is logarithmic, determined by the whole interval $T/v_F<q<1/d$. In order for the results to be valid for some real sample with a finite in-plane size, this size should be much larger than the corresponding convergence scale $1/q$ specified here. The convergence scale of the frequency integral is $\omega\sim T$ in all low temperature cases, while in the high temperature cases it is $v_F\sqrt{\kappa/d}$ in the plasmonic case, $D\kappa/d$ in the diffusive case, and $v_F/d$ in the clean case. 

The specified $q$ and $\omega$ scales also determine the condition for the validity of the Coulomb limit, which is $q\gg\omega/c$. In both clean regimes this condition reduces to $v_F\ll{c}$, which is always true. In both plasmonic regimes, the condition $q\gg\omega/c$ translates into a condition on the antisymmetric plasmon velocity, $v_F\sqrt{\kappa{d}}\ll{c}$. In the high-temperature diffusive regime, the condition is $v_F\kappa\ell\ll{c}$. Finally, in the low-temperature diffusive regime, the condition reads
\begin{equation}\label{eqn:ldCoulomb}
\frac{Td}c\ll\frac{c}{v_F}\,\frac1{\kappa\ell}.
\end{equation}
Note that none of these cases corresponds to comparing the separation $d$ to the thermal wavelength $c/T$.

Plasmon contributions dominate in a parametrically wide region of the parameter space. Crucially, this behaviour is captured by neither of the limiting expressions (\ref{eqn:Piclean}) nor (\ref{eqn:Pidiff}), but by the leading term of the small~$q$ expansion of the full expression~(\ref{eqn:polarizability}). This is equivalent to using the Drude expression for the optical conductivity, $\sigma(\omega)=e^2\nu{D}/(1-i\omega\tau)$, that is, neglecting the spatial dispersion. We find that the spatial dispersion can also be neglected to describe the diffusive contribution. It becomes important only in the clean region where $J(T)$ is dominated by the hatched area in Fig.~\ref{fig:qw-plane}. In all three high-temperature regions, one can approximate the Bose distribution as $\mathcal{N}(\omega)\approx{T}/\omega$, so the density fluctuations can be treated classically and the resulting $J(T)\propto{T}$.

Expressions (\ref{eqn:results}) were obtained for two identical metallic layers. If they are different, but the material parameters $\nu,v_F$ and $\tau$ are of the same order, our expressions are still valid as order-of-magnitude estimates. In particular, this applies to the plasmon contribution: the plasmons remain strongly hybridised even when the bare plasmonic dispersions of each layer do not match exactly. The case when the two materials are strongly different, is beyond the scope of the present paper.

\subsection{Generality of the results} 
\label{ssec:applicability}

The above results were derived for a very specific model of a 2D metal, described in Sec.~\ref{ssec:model}. We now discuss how sensitive these results are to the details of the model.

One assumption used in derivation of Eq.~(\ref{eqn:polarizability}) is that the electron scattering is dominated by short-range impurities. If it is relaxed, Eq.~(\ref{eqn:polarizability}) is not valid quantitatively (see Appendix~\ref{app:response}). Note, however, that in the clean regimes (lc, hc), the result is not sensitive to the electron scattering at all, so expressions (\ref{eqn:Jlc}) and (\ref{eqn:Jhc}) remain quantitatively valid. In the other four regimes (ld, hd, lp, hp, and crossovers between them) it is sufficient to use only the local Drude conductivity $\sigma(\omega)$, which is very general. Thus, all information about the electron scattering, needed to describe these regimes, is encoded in the momentum relaxation time~$\tau$, and expressions (\ref{eqn:Jlp})--(\ref{eqn:Jhd}) remain valid for any disorder (weak enough, not to induce Anderson localisation effects). 

If the electron momentum relaxation is due to some inelastic scattering mechanism (such as electron-phonon), the relaxation time acquires a temperature dependence. Then, the screened Coulomb interaction depends on both temperatures through the respective polarisation operators, so the heat current can no longer be written as $J(T_1,T_2)=J(T_1)-J(T_2)$. Even if the layers are made of the same material, when kept at different temperatures, they should be treated as different because their inelastic scattering times are different. If the two temperatures are of the same order, our expressions can still be used as order-of-magnitude estimates; the situation when they are strongly different is beyond the scope of this paper.

Another strong assumption behind Eq.~(\ref{eqn:polarizability}) is the strictly 2D character of the electron motion, valid for atomically thin materials. For a metallic slab whose thickness~$h$ exceeds a few Fermi wavelengths, several electronic transverse modes contribute to the density response, making Eq.~(\ref{eqn:polarizability}) invalid even in the clean limit. However, in electrodynamics, the conditions for a material slab to be described as an infinitely thin layer with \emph{some} density response function $\Pi(\mathbf{q},\omega)$, are much weaker. Namely, (i)~the normal component of the electric field should not penetrate inside the slab, since the description in terms of a 2D density response function implies that the electrons respond only to the in-plane component, and (ii)~the in-plane component should be approximately constant over the slab thickness. The first condition is usually satisfied in metals at frequencies below the bulk plasma frequency~$\omega_p$ (typically, several eV), when the normal electric field component is  screened on the length scale of the bulk screening radius (typically, on the atomic scale). The second condition requires the layer thickness~$h$ to be smaller compared to both $1/q$ and the skin depth $\delta(\omega)$ in the corresponding regime of the skin effect~\cite{Landau1981}. The relevant values of $q$ and $\omega$ are determined by the convergence scales of the corresponding integrals [as specified after Eqs.~(\ref{eqn:results})]. Note that in all cases the 2D response function must satisfy $\Pi(q\ll1/h,\omega=0)=-\nu$, where $\nu=\nu_{3D}h$ is determined by the bulk density of states per unit volume,~$\nu_{3D}$. This defines $\kappa=2\pi{e}^2\nu$, as before.
Since thicker layers imply larger $\nu$ and $\kappa$, the convergence scale of $q$ may become quite small in some of the discussed regimes, so the Coulomb limit condition $q\gg{\omega}/c$ may become violated for $h$~not small enough.

\section{Derivation of the results}
\label{sec:results}

For identical layers, Eq.~(\ref{eqn:transmission}) can be written in a slightly simpler form:
\begin{equation}\label{eqn:transmissionPiPi}
\frac{\mathcal{T}}2=\left|\frac{\Im\Pi\,{v}_qe^{-qd}}{[1-\Pi{v}_q(1-e^{-qd})][1-\Pi{v}_q(1+e^{-qd})]}\right|^2.
\end{equation}
Below, we analyse the contributions from the three regions of the $(q,\omega)$~plane in Fig.~\ref{fig:qw-plane}: the vicinity of the plasmon dispersions, the diffusive region, and the clean region. In each region, we separate two temperature regimes. We see that in the high-temperature regimes, the integral~(\ref{eqn:heat}) is dominated by a certain frequency scale, different in each regime, but always determined by $q\sim1/d$ and being much lower than the temperature. Then, the thermal cutoff plays no role, and one can approximate the Bose distribution $\mathcal{N}(\omega)\approx{T}/\omega$ [that is why $J(T)\propto{T}$ in Eqs.~(\ref{eqn:Jhc}), (\ref{eqn:Jhp}) and (\ref{eqn:Jhd})]. In the low-temperature regimes, it is the $e^{-qd}$ cutoff which is ineffective, so the frequency integral is determined by $\omega\sim{T}$, while the $q$~integral converges on a scale much smaller than $1/d$. Then one can approximate $e^{-qd}\approx{1}$ in the numerator of Eq.~(\ref{eqn:transmissionPiPi}) and $1+e^{-qd}\approx2$, $1-e^{-qd}\approx{q}d$ in the denominator.

\subsection{Plasmon contribution}

As discussed earlier, the plasmon contribution comes from the region $\omega>1/\tau,v_Fq$, since otherwise the plasmons are overdamped.
At such frequencies one can expand $\Pi(q,\omega)$ to the leading order in~$q$ and approximate
\begin{equation}\label{eqn:Pivq}
\Pi{v}_q=\frac{v_F^2\kappa{q}}{2\omega(\omega+i/\tau)},
\end{equation}
which corresponds to neglecting the spatial dispersion in the conductivity which takes the Drude form, $\sigma(\omega)=e^2\nu{D}/(1-i\omega\tau)$. (Generally, the spatial dispersion can be neglected when $q\ll\max\{\sqrt{\omega/D},\omega/v_F\}$.) The integral is then dominated by the vicinities of the two plasmon dispersions, where one of the factors in the denominator of Eq.~(\ref{eqn:transmissionPiPi}) is small. The plasmon contribution exists only if the plasmon frequencies $\omega_\pm$ at $q\sim1/d$ (when the spatial cutoff becomes effective) 
exceed $1/\tau$. This gives a condition $d\ll\kappa\ell^2$ (the vertical line $x=\eta^2$ in Fig.~\ref{fig:domregions}).

Let us first consider the temperature interval $1/\tau\ll T\ll{v}_F\sqrt{\kappa/d}$ (the two inequalities are consistent when $d\ll\kappa\ell^2$), where the thermal cutoff plays first leaving the spatial cutoff $e^{-qd}$ ineffective. Then one can expand $e^{-qd}$ and perform the $q$~integration first, approximating the integrand by a Lorentzian in the vicinity of each pole. The remaining frequency integral can be calculated exactly:
\begin{align}
J_{\rm lp}(T)={}&{}\int_0^\infty\frac{\omega\,d\omega/\pi}{e^{\omega/T}-1}
\frac\omega{4\tau{v}_F^2\kappa^2}\left(\frac{\omega^2}{v_F^2}+\frac\kappa{d}\right)
\nonumber\\
={}&{}\frac{T^3}{2\pi\tau{v}_F^2\kappa{d}}\left[\zeta(3)+12\,\zeta(5)\,\frac{T^2d}{v_F^2\kappa}\right].
\end{align}
Here $\zeta(x)$ is the Riemann zeta function, and the first (second) term in the square brackets comes from the antisymmetric (symmetric) plasmons, respectively. In the considered region $1/\tau\ll T\ll{v}_F\sqrt{\kappa/d}$ the symmetric contribution is always small compared to the antisymmetric one.

At high temperatures, $T\gg{v}_F\sqrt{\kappa/d}$, the integral is determined by the spatial factor $e^{-qd}$, while the thermal cutoff is ineffective so the Bose distribution $\mathcal{N}(\omega)\approx{T}/\omega$. Taking expression~(\ref{eqn:Pivq}), Eq.~(\ref{eqn:transmissionPiPi}) can also be written in the form suitable for integration over~$\omega$:
\begin{equation}
\mathcal{T}(q,\omega)=\frac{\omega^2}{2\tau^2}\,\frac{(\omega_+^2-\omega_-^2)^2}%
{|\omega(\omega+i/\tau)-\omega_+^2|^2|\omega(\omega+i/\tau)-\omega_-^2|^2}.
\end{equation}
Note that one cannot just do two separate Lorentzian integrals because the separation $\omega_+^2-\omega_-^2$ becomes exponentially small at $q\gtrsim1/d$.
Fortunately, the $\omega$~integral can be calculated exactly:
\begin{equation}
\int\limits_0^\infty\frac{d\omega}\pi\,\mathcal{T}(q,\omega)=
\frac{1}{2\tau}\,\frac{(\omega_+^2-\omega_-^2)^2}{(\omega_+^2-\omega_-^2)^2+2(\omega_+^2+\omega_-^2)/\tau^2}.
\end{equation}
Then the $q$~integral reads as
\begin{equation}\label{eqn:Jhpint}
J_{\rm hp}(T)=\frac{T}{\tau}\int_0^\infty\frac{q\,dq}{4\pi}\,
\frac{\ell^2\kappa{q}}{\ell^2\kappa{q}+2e^{2qd}},
\end{equation}
leading to Eq.~(\ref{eqn:Jhp}) with the function $\mathcal{L}(x)$ defined as
\begin{equation}\label{eqn:Ldef}
\mathcal{L}(x)\equiv\int_0^\infty\frac{u^2\,du}{u+e^u/x}.
\end{equation}

\subsection{Diffusive contribution}
\label{ssec:diffusive}

Let us focus on the contribution from the shaded region~in Fig.~\ref{fig:qw-plane}: $q\ll1/\ell$, $\omega\ll1/\tau$. Then one can use expression~(\ref{eqn:Pidiff}) for $\Pi(q,\omega)$, and since we are interested in $q\lesssim1/d\ll\kappa$, we have $Dq^2\ll{D}\kappa{q}(1\pm{e}^{-qd})/2$, which is again equivalent to neglecting the spatial dispersion in the conductivity. Thus, we can write 
\begin{equation}\label{eqn:transmissionDD}
\mathcal{T}\approx\frac{2\omega^2(D\kappa{q})^2e^{-2qd}}%
{[\omega^2+(D\kappa{q})^2(1+e^{-qd})^2][\omega^2+(D\kappa{q})^2(1-e^{-qd})^2]}.
\end{equation}
At low frequencies, when the momentum integral should converge on some scale $q\ll1/d$, the two factors in the denominator are strongly different, and it is the second factor that determines the convergence scale $q\sim\sqrt{\omega/(D\kappa{d})}$. When $\omega\sim{T}\ll{D}\kappa/d$ this scale is indeed much smaller than $1/d$, so we expand $e^{-qd}$, integrate over~$q$, then over $\omega$, and arrive at Eq.~(\ref{eqn:Jld}). Moreover, the convergence scale $\sqrt{T/(D\kappa{d})}\ll1/\ell$ provided that $T\ll\kappa{d}/\tau$. Thus, since we always assume $\kappa{d}\gg{1}$, Eq.~(\ref{eqn:Jld}) is valid for the diffusive contribution everywhere below the horizontal line $y\sim\eta$ ($T\sim1/\tau$) in Fig.~\ref{fig:domregions}.

For $T\gg{D}\kappa/d$, the $q$ integral is dominated by $q\sim1/d$. Then the typical frequency scale of Eq.~(\ref{eqn:transmissionDD}) is $\omega\sim{D}\kappa/d$, so for $T\gg{D}\kappa/d$ the thermal cutoff is ineffective. Then we approximate $\mathcal{N}(\omega)\approx{T}/\omega$, straightforwardly integrate over~$\omega$, then over~$q$, and arrive at Eq.~(\ref{eqn:Jhd}). Note that  Eq.~(\ref{eqn:Jhd}) is valid even at $T\gg{1}/\tau$ provided that the convergence scale $\omega\sim{D}\kappa/d\ll1/\tau$, that is, to the left of the vertical line $x=\eta^2$ in Fig.~\ref{fig:domregions}. As we have seen, to the right of this line the plasmon contribution becomes important. 

\subsection{Clean contribution}
\label{sec:clean}

For $T\gg1/\tau$, $1/d\gg1/\ell$, one should take into account the contribution from the hatched area in Fig.~\ref{fig:qw-plane}. Here one can take the limit $\tau\to\infty$ and use expression~(\ref{eqn:Piclean}) for $\Pi(q,\omega)$. Then, in the integration region $\omega<v_Fq$, $\Re\Pi(q,\omega){v}_q=-\kappa/q$, so one can neglect unity in both factors in the denominator of Eq.~(\ref{eqn:transmissionPiPi}) and write
\begin{equation}
\mathcal{T}\approx\frac{2}{v_q^2}\left|\frac{\Im\Pi}{\Pi^2}\right|^2
\frac{e^{-2qd}}{(1-e^{-2qd})^2}
=\frac{\omega^2(v_F^2q^2-\omega^2)}{2v_F^4\kappa^2q^2\sinh^2qd}.
\end{equation}
For $T\gg{v}_F/d$ we approximate $\mathcal{N}(\omega)\approx{T}/\omega$, straightforwardly integrate over~$\omega$ between $0$ and $v_Fq$, then integrate over~$q$, and arrive at Eq.~(\ref{eqn:Jhc}).
For $T\ll{v}_F/d$, in most of the integration region we have $\omega\sim{T}\ll{v}_Fq$, so the upper limit $\omega<v_Fq$ is not important except for the narrow region $q\sim{T}/v_F$ which determines the lower cutoff of the logarithmic $q$~integral:
\begin{equation}
J_{\rm lc}(T)=\frac{1}{4\pi^2}\int\limits_0^\infty\frac{\omega^3\,d\omega}{e^{\omega/T}-1}
\int\limits_{\sim{T}/v_F}^\infty\frac{q\,dq}{\sinh^2qd},
\end{equation}
leading to Eq.~(\ref{eqn:Jlc}).

In the region $T\gg1/\tau$, $1/d\gg1/\ell$, the clean contribution and the plasmon contribution both exist and should be added up, since they come from two distinct regions in the $(q,\omega)$~plane. Thus, to determine the dominant asymptotics, one can combine the four expressions (\ref{eqn:Jlc})--(\ref{eqn:Jhp}) as $J=\max\{\min\{J_{\rm lp},J_{\rm hp}\},\min\{J_{\rm lc},J_{\rm hc}\}\}$, which results in the complicated shape of the boundary between the clean and plasmonic regions in Fig.~\ref{fig:qw-plane}.


\begin{figure}
\includegraphics[width=0.45\textwidth]{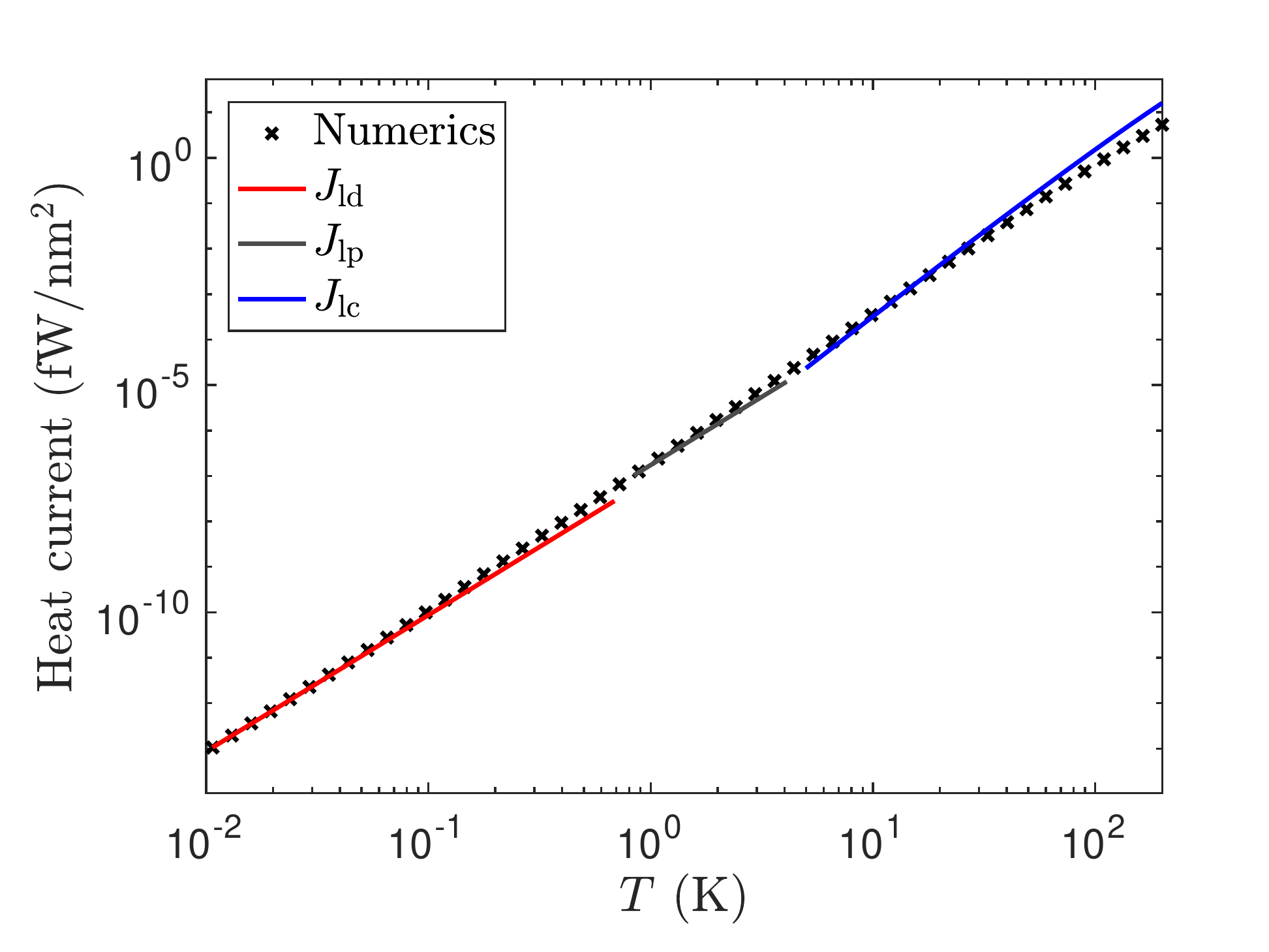}
\includegraphics[width=0.45\textwidth]{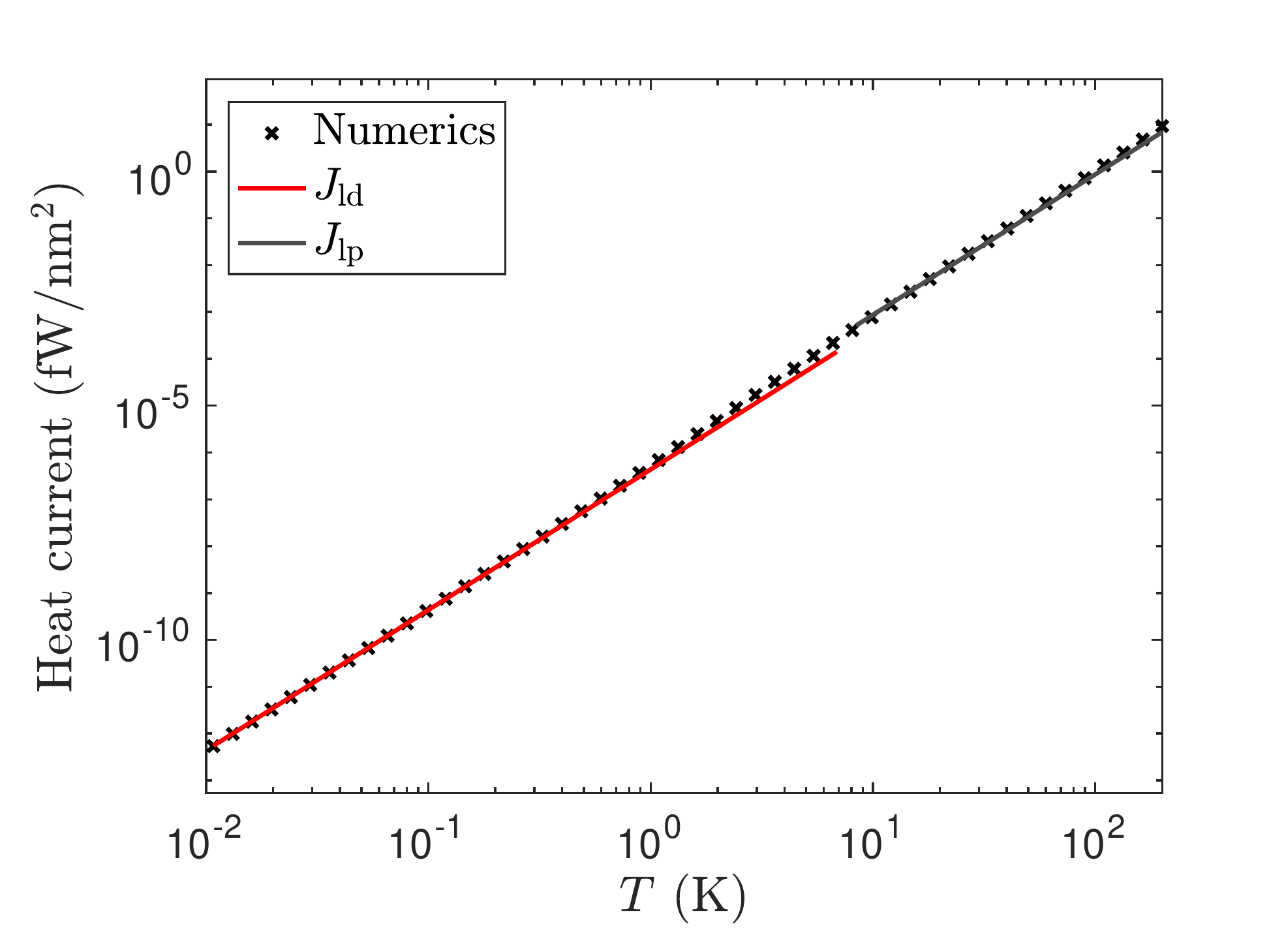}
\caption{Power per unit area as a function of temperature, $J(T)$, for $\nu=0.29\:\mbox{eV}^{-1}\mbox{nm}^{-2}$, $v_F=10^6\:\mbox{m/s}$,
and $\ell=10\:\mu\mathrm{m}$, $d=10\:\mbox{nm}$ (upper panel), 
$\ell=1\:\mu\mathrm{m}$, $d=20\:\mbox{nm}$ (lower panel), 
characteristic of two graphene monolayers with Fermi energy $\epsilon_F=0.2\:\mbox{eV}$, suspended in vacuum (the 2D screening length $\kappa^{-1}=0.4\:\mbox{nm}$, the Fermi momentum $p_F=0.3\:\mbox{nm}^{-1}$).
The black crosses show numerical results, while the red, black and blue solid lines represent Eqs. (\ref{eqn:Jld}), (\ref{eqn:Jlp}) and (\ref{eqn:Jlc}), respectively.
}
\label{fig:graphene}
\end{figure}

\section{Numerical verification}
\label{sec:numerics}

In order to illustrate the various behaviours and crossovers indicated in Fig.~\ref{fig:domregions}, and to verify our asymptotic expressions~(\ref{eqn:results}), we evaluate numerically the integral in Eq.~(\ref{eqn:heat}) using the full response function~(\ref{eqn:polarizability}).

To describe clean and plasmonic regimes, we take parameters typical for doped graphene. If the temperature is well below the Fermi energy $\epsilon_F$ (counted from the Dirac point) the electrons in graphene can be viewed as a conventional 2D electron gas with the density of states per unit area $\nu=2|\epsilon_F|/(\pi{v}_F^2)$, including the valley and spin degeneracies. In Fig.~\ref{fig:graphene}, we plot $J(T)$ for two sets of parameters corresponding to $1/(\kappa{d})\approx\eta^{1/3}=0.034$, and $1/(\kappa{d})\approx\eta^{1/2}$ [we remind that $\eta\equiv1/(\kappa\ell)$].

To study the diffusive crossover for realistic materials, we introduce dielectric screening.
Equation~(\ref{eqn:V12}), written for metallic layers surrounded by vacuum, can be generalised to the situation when the layers are embedded in a dielectric medium. This generalisation is particularly simple when the medium is characterised by a uniaxially anisotropic dielectric constant, $\varepsilon_\|$ in the plane parallel to the layers, and $\varepsilon_\perp$ along the $z$ direction, perpendicular to the layers. The solution of the Poisson equation in such a medium gives the 2D Coulomb potential at a distance~$z$ from a charged layer:
\begin{equation}
v_\mathbf{q}(z)=\frac{2\pi{e}^2}{\sqrt{\varepsilon_\|\varepsilon_\perp}\,q}\,
e^{-\sqrt{\varepsilon_\|/\varepsilon_\perp}\,q|z|}.
\end{equation}
Thus, in all expressions~(\ref{eqn:results}) it is sufficient to rescale 
\begin{equation}
\kappa\to\frac{\kappa}{\sqrt{\varepsilon_\|\varepsilon_\perp}},\quad
d\to\sqrt{\varepsilon_\|/\varepsilon_\perp}\,d.
\end{equation}

\begin{figure}
\includegraphics[width=0.45\textwidth]{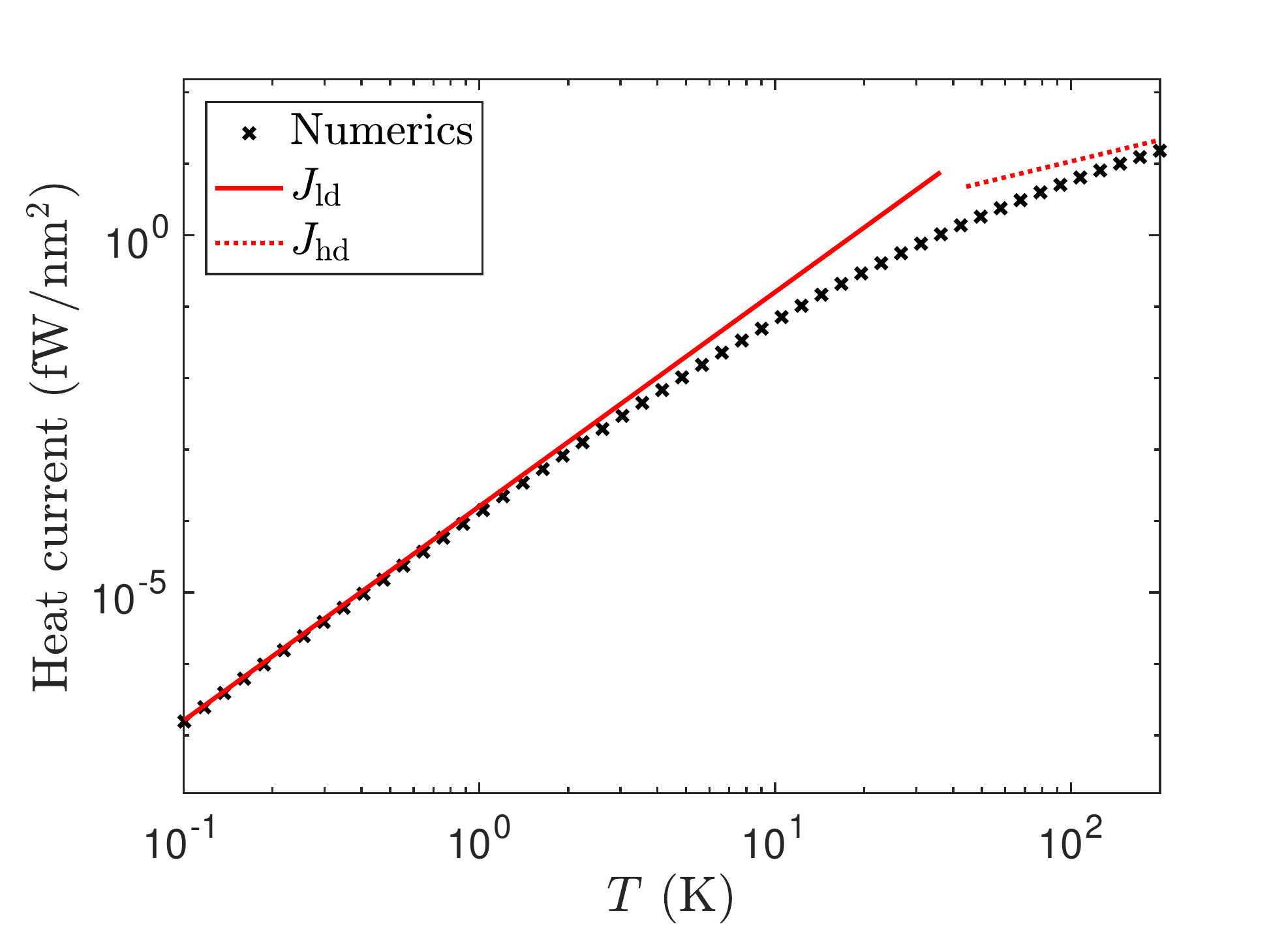}
\caption{Power per unit area as a function of temperature, $J(T)$, for $\nu=2.1\:\mbox{eV}^{-1}\mbox{nm}^{-2}$, $v_F=0.94\times10^5\:\mbox{m/s}$,
and $\ell=2\:\mu\mathrm{m}$, $d=100\:\mbox{nm}$, 
characteristic of two WSe$_2$ monolayers with Fermi energy $\epsilon_F=50\:\mbox{meV}$, embedded in boron nitride with $\epsilon_\|=7$, $\epsilon_\perp=5$ \cite{Geick1966} (the 2D screening length $\kappa^{-1}=0.3\:\mbox{nm}$, the Fermi momentum $p_F=1.6\:\mbox{nm}^{-1}$).
The black crosses show the numerical integration results, while the solid and dashed red lines represent expressions (\ref{eqn:Jld}) and (\ref{eqn:Jhd}), respectively.
}
\label{fig:WSe2}
\end{figure}

In Fig.~\ref{fig:WSe2} we show the crossover between low- and high-temperature diffusive asymptotics (\ref{eqn:Jld}) and (\ref{eqn:Jhd}) for two hole-doped tungsten diselenide monolayers embedded in boron nitride. The valence band of WSe$_2$ is parabolic with the hole effective mass $m_h$ being about half of the free electron mass. The spin degeneracy is lifted by a strong spin-orbit coupling, so only valley degeneracy remains, and the density of states per unit area is $\nu=m_h/\pi$. We take  $\epsilon_F=50\:\mbox{meV}$ and a very short $\ell=2\:\mbox{nm}$, still consistent with $\kappa\ell\gg1$, $p_F\ell\gg1$. The taken separation $d=100\:\mbox{nm}$ corresponds to $1/(\kappa{d})=0.006$, well below $\eta^2=0.1$, and hence to the diffusive region.

We are not showing the high-temperature clean and plasmonic regimes; for realistic material parameters, they correspond to temperatures so high that the assumptions behind our model (degenerate Fermi gas, near-field Coulomb regime) are no longer valid. However, we checked numerically the validity of the asymptotic expressions (\ref{eqn:Jhc}) and (\ref{eqn:Jhp}) for $J_\mathrm{hc}(T)$ and $J_\mathrm{hp}(T)$.

\section{Comparison to circuit theory}
\label{sec:circuit}

Often, complicated structures can be described in terms of effective electric circuits  made of lumped elements (capacitors, inductors, and resistors). In this approach, all details of the structure's geometry are hidden inside the effective circuit parameters, resulting in a much simpler description (provided that such reduction is valid). The theory of heat transfer in electric circuits was developed in Ref.~\cite{Pascal2011}. In the circuit analog of the fluctuational electrodynamics, dissipative circuit elements represent thermal baths and provide thermal voltage fluctuations (Johnson-Nyquist noise), while reactive elements mediate the electromagnetic interactions, resulting in energy exchange between the baths. We check now whether such a circuit approach can be applied to the heat transfer between two metallic layers.

\begin{figure}
\centering
\includegraphics[width=0.35\textwidth]{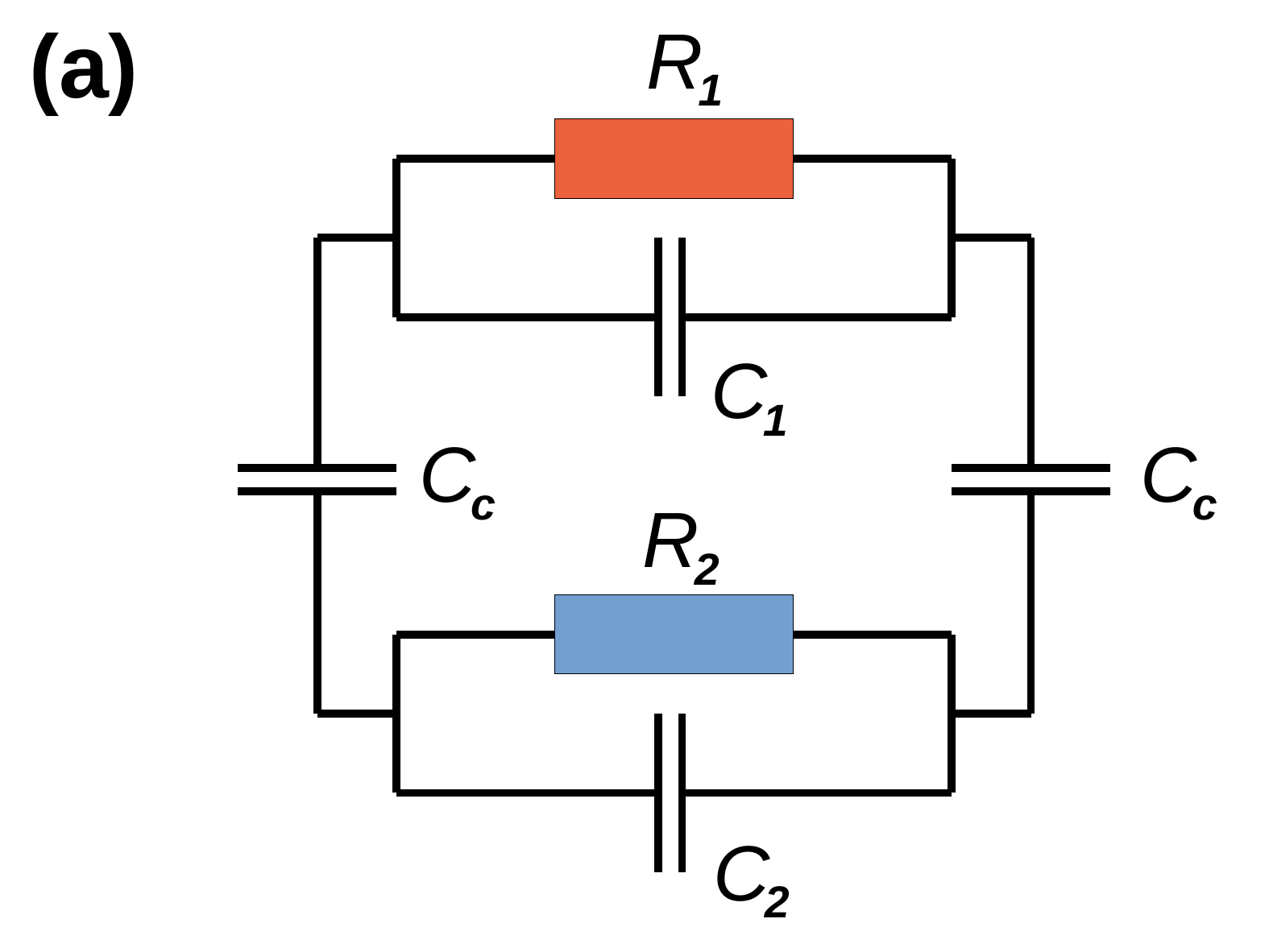}
\includegraphics[width=0.48\textwidth]{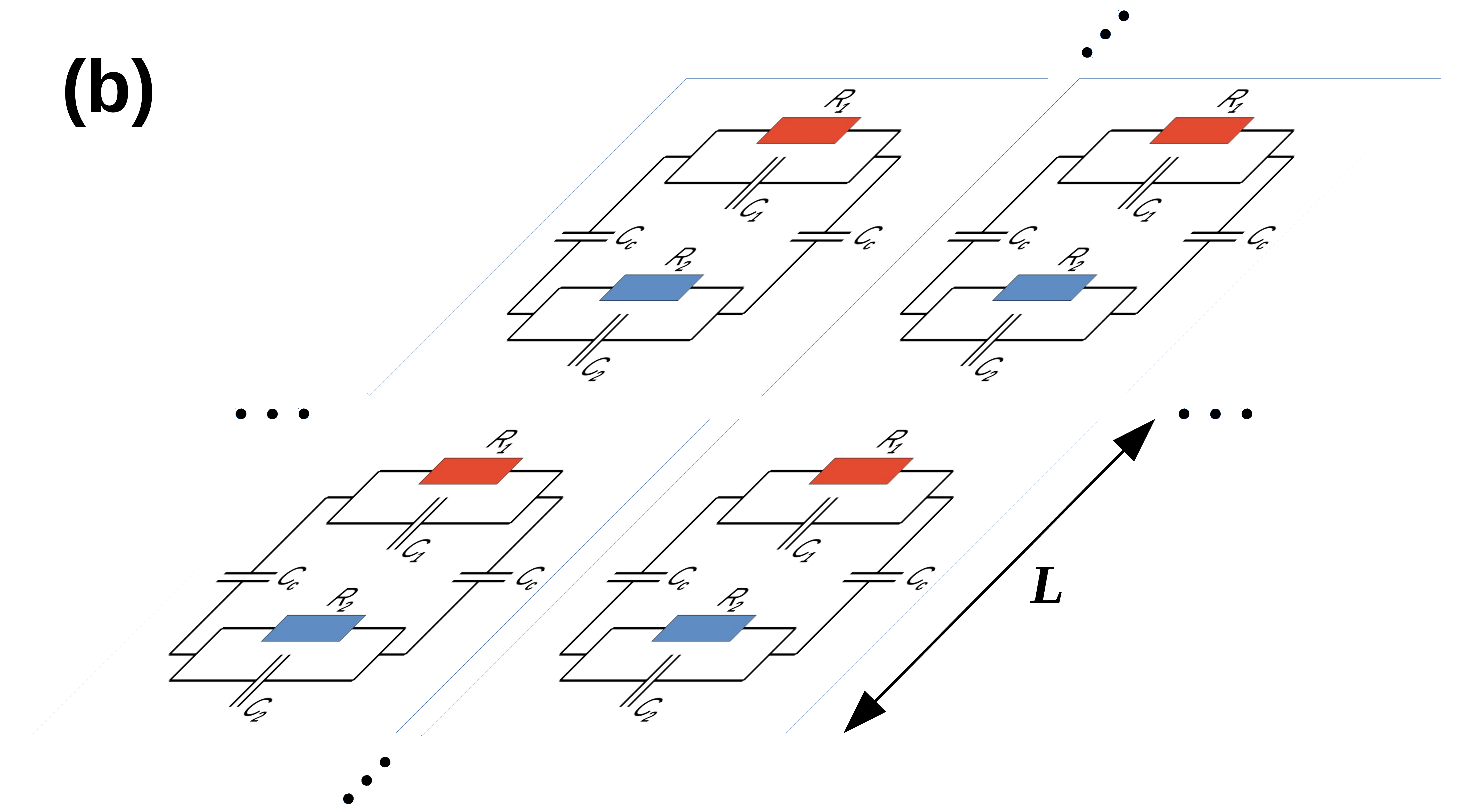}
\caption{Circuit representation of two parallel metallic layers. (a)~The elementary $RC$ circuit described by Eq.~(\ref{eqn:Tcircuit12}). (b)~The 2D system represented as a tiling of the elementary circuits, each one corresponding to a region of size~$L$.}
\label{fig:circuit}
\end{figure}

Let us focus on the diffusive regime. As we have seen, the dynamics of density excitations is overdamped in this regime, so it is natural to consider a circuit made of resistors and capacitors only, such as shown in Fig.~\ref{fig:circuit}(a). Indeed, the electronic excitations in each layer constitute a dissipative bath analogous to a resistor. To mimic charge oscillations within each layer, the resistor should be shunted by a capacitor. The Coulomb interaction between the layers resembles that between the plates of a capacitor, so the two $RC$ contours are connected by two coupling capacitors~$C_c$.


For the circuit in Fig.~\ref{fig:circuit}(a),
the power transferred from the first to the second resistor is given by~\cite{Pascal2011}
\begin{align}
&P(T_1,T_2) = \int_0^\infty \frac{d\omega}\pi\,
\omega\left[\mathcal{N}_1(\omega)-\mathcal{N}_2(\omega)\right]\mathcal{T}(\omega),\\
&\mathcal{T}(\omega)=\frac{2\,\Re{Z}_1(\omega)\,\Re{Z}_2(\omega)}{|{Z}_1(\omega)+{Z}_2(\omega)-2/(i\omega{C_c})|^2},\label{eqn:Tcircuit12}
\end{align}
where $Z_{1,2}(\omega)=(1/R_{1,2}-i\omega{C}_{1,2})^{-1}$ is the impedance of each $RC$ contour. As before, for simplicity we assume the two subsystems to be identical: $R_1=R_2=R$, $C_1=C_2=C$.
Writing the transmission as
\begin{equation}\label{eqn:Tcircuit}
\mathcal{T}(\omega)=\frac{\omega^2({R}C_c)^2}{2[1+\omega^2({R}C)^2][1+\omega^2R^2(C+C_c)^2]},
\end{equation}
we obtain the following asymptotic expressions for the transferred power $P(T_1,T_2)=P(T_1)-P(T_2)$:
\begin{subequations}\label{eqn:power}\begin{align}
&P(T)=\frac{C_c^2}{4(C+C_c)(2C+C_c)}\,\frac{T}{RC},\quad
T\gg\frac{1}{RC},\label{eqn:Ph}\\
&P(T)=\frac\pi{12}\,T^2,\quad
\frac{1}{R(C+C_c)}\ll T\ll\frac{1}{RC},\label{eqn:Pu}\\
&P(T)=\frac{\pi^3}{30}\,T^4(RC_c)^2,\quad T\ll\frac{1}{R(C+C_c)}.\label{eqn:Pl}
\end{align}\end{subequations}
The intermediate ``universal'' regime, where the power depends only on the temperature, but not on the circuit, is present only when $C\ll{C}_c$.

To relate results~(\ref{eqn:power}) to those of Sec.~\ref{sec:results}, it is important to realise that while Eqs.~(\ref{eqn:power}) give the full transferred power, Eqs.~(\ref{eqn:results}) give the power \emph{per unit area}. To make a meaningful comparison we must therefore invoke a length scale~$L$, such that the infinite sample can be divided into squares of size~$L$. Then Eqs.~(\ref{eqn:power}) describe the power transferred in each square, and the contributions of different squares can be added up independently, as schematically shown in Fig.~\ref{fig:circuit}(b).
Thus, the relevant length scale should be associated with the typical convergence scale of the $q$~integral in Eq.~(\ref{eqn:heat}).


In the diffusive regime, it is natural to associate the resistance $R$ to the resistance per square $1/\sigma$ of each metallic layer, $R\sim1/\sigma=2\pi/(\kappa{D})$. The coupling capacitance is associated to the geometric capacitance between the two layers, $C_c\sim{L}^2/d$, where the in-plane length scale~$L\gtrsim{d}$ must be invoked because the capacitance is proportional to the area. The capacitance~$C$ should be associated to an intrinsic property of each layer, such that $RC$ corresponds to the characteristic relaxation time of charge density modulations. In a 2D system, this time depends on the in-plane length scale $L$ and is given by $L/\sigma$. This gives $C\sim{L}$. Recalling the convergence scales of the $q$~integral in Sec.~\ref{ssec:diffusive}, we associate $L\sim\sqrt{D\kappa{d}/T}$ and $L\sim{d}$ in the low- and high-temperature diffusive regimes, respectively. Then Eqs.~(\ref{eqn:Pl}) and~(\ref{eqn:Ph}) match Eqs.~(\ref{eqn:Jld}) and~(\ref{eqn:Jhd}) at low and high temperatures, respectively. Expression~(\ref{eqn:Pu}) does not correspond to any parametric region because at $T\gg{D}\kappa/d$ the two capacitances become of the same order.

To summarise, while the proposed effective $RC$ circuit does capture the qualitative picture of the heat transfer in the diffusive regime, one cannot completely disregard the 2D geometry of the system. This geometry manifests itself in the appearance of the length scale~$L$, which must supplement the circuit picture in order to reproduce the temperature dependence of the heat current. Moreover, this length scale is temperature-dependent, so one cannot represent a given system by a given array of elementary circuits in the whole temperature range. This strongly limits the usefulness of the circuit analogy. For this reason, we do not consider the other two regimes (plasmonic and clean), whose modelling would require a more complicated circuit; this would have to include also inductors in order to mimic the physics of weakly damped density excitations.

\section{Discussion of experiments}
\label{sec:experiments}

Let us now discuss some relevant experiments from the viewpoint of the theory presented in this paper. In the literature we are aware of, experiments in the planar geometry are not very numerous; apparently, a tip-to-plane or sphere-to-plane configurations are easier to realise in a controllable way.

In a recent experiment~\cite{Yang2018}, two monolayer graphene sheets were placed on insulating silicon (the dielectric constant $\varepsilon=11.7$) and separated by a vacuum gap of width $d=430\:\mbox{nm}$. The experiment was performed around room temperature, and the linear thermal conductance per unit area $dJ/dT=30\:\mbox{W}\,\mbox{m}^{-2}\,\mbox{K}^{-1}$ was measured. Importantly, the doping level was sufficiently high, $\epsilon_F=0.27\:\mbox{eV}$, so the peculiarities of the Dirac spectrum can be safely ignored and the electrons can be treated as the usual 2D electron gas with the 2D screening length $\kappa^{-1}=2.8\:\mbox{\AA}$. The single-layer plasmon frequency at $q=1/d$ is $v_F\sqrt{\kappa/(\varepsilon+1)d)}\sim3\times10^{13}\:\mbox{s}^{-1}$, slightly below the room temperature ($T=300\:\mbox{K}$ corresponds to $\omega=3.93\times10^{13}\:\mbox{s}^{-1}$) and above $1/\tau=10^{13}\:\mbox{s}^{-1}$~\cite{Yang2018}. These conditions correspond to the crossover between the low- and high-temperature plasmon regimes. The plasmon velocity is about 20 times smaller than the speed of light, so the heat transfer is well within the Coulomb limit.

To give a theoretical estimate for the thermal conductance, one has to account for the strong dielectric contrast (vacuum vs silicon) between the two sides of each graphene layer. This amounts to replacing $v_\mathbf{q}\Pi\to v_\mathbf{q}\Pi+(1-\varepsilon)/2$ in the denominator of Eq.~(\ref{eqn:V12}). In the high-temperature plasmon regime,
one should multiply $2e^{2qd}$ in the denominator of Eq.~(\ref{eqn:Jhpint}) by
\[
\frac{[(\varepsilon+1)^2-(\varepsilon-1)^2e^{-2qd}][(\varepsilon+1)-(\varepsilon-1)e^{-2qd}]}8,
\]
so $\varepsilon$ enters only inside the logarithmic function $\mathcal{L}$. Setting $\mathcal{L}=1$ gives $dJ/dT=11\:\mbox{W}\,\mbox{m}^{-2}\,\mbox{K}^{-1}$, which agrees by order of magnitude with the experimental value.

Several works have been dedicated to transfer between finite-thickness layers of conventional metals, to be reviewed below. In each case, one should decide whether these layers should correspond to the 2D or bulk limit, according to the discussion in Sec.~\ref{ssec:applicability}. In all cases, the separation $d$ between the sheets is larger than their thickness $h$, so the criterion $h\ll1/q$ is always fulfilled. Thus, the main issue is the comparison between the layer thickness and the skin depth.

In the pioneering study in Ref.~\cite{Hargreaves1969}, heat transfer between two $h=100\:\mbox{nm}$ thick chromium plates was measured at room temperature, probing separation dependence of heat flux over scales $d=1-8\:\mu\mbox{m}$. Reference~\cite{Hargreaves1969} provides no information on electron scattering, but if we take a typical value for the mean free path $\ell_{3D}=10\:\mbox{nm}$ and use the values $\nu_{3D}=0.06\:\mbox{eV}^{-1}\mbox{\AA}^{-3}$~\cite{Rath1973}, $v_F=0.5\times10^6\:\mbox{m/s}$~\cite{Sinha1969},  we obtain the normal skin depth $\delta=50\:\mbox{nm}$ at frequency $\omega=3.93\times10^{13}\:\mbox{s}^{-1}$, corresponding to $T=300\:\mbox{K}$. Thus, such plates cannot be treated as two-dimensional.

In Ref.~\cite{Kralik2012}, heat transfer was studied for two tungsten layers $h=150\:\mbox{nm}$ thick on alumina substrate, over a wide range of separations $d=1-300\: \mu\mbox{m}$, with the cold layer held at $5\:\mbox{K}$, while the hot layer temperature was varied in the range $10-40\:\mbox{K}$.
Specifically, at small separations, the observed temperature and distance dependence was $J(T)\propto{T}^{1.5}/d^{2.5}$.
The measured dc conductivity of the material, corresponding to $4\pi\sigma_{3D}=0.6\times10^{18}\:\mbox{s}^{-1}$, was constant in the temperature range $4-77\:\mbox{K}$, identifying the static disorder as the dominant source of electron scattering with $\tau=6\:\mbox{fs}$ and $\ell_{3D}=6\:\mbox{nm}$ ($v_F=10^6\:\mbox{m/s}$ from Ref.~\cite{Gall2016}). This gives the normal skin depth corresponding to $T=40\:\mbox{K}$ as $\delta=240\:\mbox{nm}$, and even longer at lower temperatures. Thus, one could expect this experiment to be correspond to the low-temperature diffusive 2D regime, whose conditions can be written as $T\ll1/\tau$, $T\ll4\pi\sigma_{3D}h/d$.
However, if we check the Coulomb limit condition~(\ref{eqn:ldCoulomb}) estimating $\nu_{3D}=10^{47}\:\mbox{J}^{-1}\:\mbox{m}^{-3}=0.016\:\mbox{eV}^{-1}\,\mbox{\AA}^{-3}$ from $\omega_p^2=4\pi\nu_{3D}v_F^2/3$ with $\omega_p=0.97\times10^{16}\:\mbox{s}^{-1}$~\cite{Kralik2012}, we obtain $\kappa\ell=1.3\times10^5$, and the two sides of the inequality become comparable only for the smallest $d=1\:\mu\mbox{m}$ and $T=10\:\mbox{K}$.
Thus, the whole experimental curve corresponds to the regime when retardation is important, so it is natural that the experimental result is not consistent with $J_\mathrm{ld}(T)\propto{T}^3/d$.

Linear response thermal conductance $dJ(T)/dT$ at room temperature was measured in Ref.~\cite{Song2016} for two gold layers $h=100\:\mbox{nm}$ thick with separations $60\:\mbox{nm}<d<10\:\mu\mbox{m}$. At small separations a dependence $1/d^\alpha$ with $\alpha$ between 2 and 3 was observed. These results were successfully interpreted using the Drude model parameters $\omega_p=0.6\times10^{16}\:\mbox{s}^{-1}$ and $\tau=6\:\mbox{fs}$~\cite{Ordal1985}, which give the skin depth $\delta=50\:\mbox{nm}$. 
Estimating $\nu_{3D}$ from $\omega_p$ and $v_F=1.4\times10^6\:\mbox{m/s}$~\cite{Gall2016}, we again obtain that inequality~(\ref{eqn:ldCoulomb}) is not satisfied, although for the smallest separation the two sides become comparable.

Finally, a recent preprint~\cite{Sabbaghi2019} reports a series of measurements of heat transfer between two aluminium films on silicon, separated by a vacuum gap $d=215\:\mbox{nm}$. The film thicknesses $h$ were varied in the interval from $13-79\:\mbox{nm}$.
The experiment was performed around room temperature, and for the thinnest films with $h=13\:\mbox{nm}$ the linear thermal conductance per unit area $dJ/dT=60\:\mbox{W}\,\mbox{m}^{-2}\,\mbox{K}^{-1}$ was measured. For thicker layers, the thermal conductance was slightly smaller (about $50\:\mbox{W}\,\mbox{m}^{-2}\,\mbox{K}^{-1}$ for $d=79\:\mbox{nm}$). No information on electron scattering was given in Ref.~\cite{Sabbaghi2019}, but the 13~nm thick sample was thinner than the smallest possible skin depth $c/\omega_p=30\:\mbox{nm}$ (with $\omega_p=0.97\times10^{16}\:\mbox{s}^{-1}$~\cite{Ordal1985}), so one would expect it to be in the 2D limit. From $\nu_{3D}=1.45\times10^{47}\:\mbox{J}^{-1}\:\mbox{m}^{-3}=0.023\:\mbox{eV}^{-1}\,\mbox{\AA}^{-3}$ \cite{Court2008}, we obtain the 2D screening length $1/\kappa=3.6\times10^{-3}\:\mbox{\AA}$. If we take a typical value $\ell=10\:\mbox{nm}$,
we nominally find the system to be in the low-temperature diffusive limit. Still, with  $v_F=1.6\times10^6\:\mbox{m/s}$~\cite{Gall2016}, inequality~(\ref{eqn:ldCoulomb}) is not satisfied with the left-hand side a few times larger than the right-hand side.

To summarise, metallic layers studied in experiments may have thicknesses a few times larger or a few times smaller than the skin depth, but the inequality is never very strong. 
More stringent is condition~(\ref{eqn:ldCoulomb}) which enables one to neglect retardation effects and use the Coulomb limit. For typical metals, it requires temperatures below 100$\:\mbox{K}$, while $h$ and $d$ should be in the range of tens of nanometers.

\section{Conclusions}

We have studied the problem of heat transfer between two thin parallel metallic layers, mediated by the Coulomb interaction. Using a simple model for a 2D electron gas subject to scattering on short-range impurities, we described the crossover between clean and diffusive limits and showed that strongly coupled surface plasmons dominate the heat transfer in a parametrically wide region at sufficiently high temperatures, but their contribution is suppressed in both the clean and diffusive limits. We also clarified the role of the spatial dispersion in the optical conductivity, which turns out to be important only in the clean limit. In all other regimes, the effect of disorder is correctly captured by the relaxation time in the Drude conductivity.

We have shown that in the diffusive limit, the heat transfer is quantitatively similar to that in an effective $RC$ circuit. However, for this analogy to be meaningful, one must specify a length scale. This length scale should correspond to the size of regions within the infinite 2D sample where the transfer occurs independently. In other words, each region can be described by a separate circuit, and contributions from different regions add up. This length scale must be determined from the microscopic theory
and turns out to be temperature-dependent. This greatly limits the usefulness of the circuit analogy, especially when the two temperatures are strongly different.

Comparing the theory, developed here, to experimental data, available in the literature, we found that the theory qualitatively agrees with the experiment on heat transfer between two graphene sheets, which falls into the crossover between the low- and high-temperature plasmonic regimes. 
For finite-thickness layers of conventional metals, we find that in all experiments that we are aware of, the parameters do not fulfil the stringent condition to be in the Coulomb limit when retardation can be neglected; still, this limit is realistic for nanometer-sized structures at low temperatures.

\acknowledgements
This project received funding from the European Union's Horizon 2020 research and innovation programme under the Marie Sk{\l}odowska-Curie Grant Agreement No. 766025.

\appendix

\section{Density response function from the Boltzmann equation}
\label{app:response}

We assume the 2D electron gas to be in the good metallic regime, when the mean free path~$\ell$ is much larger than the Fermi wavelength, so we may neglect localization effects. In this limit, and for perturbations smooth on the scale of the Fermi wavelength, the electron dynamics can be described by the semiclassical Boltzmann kinetic equation~\cite{Abrikosov1988}. The electron distribution function $f_\mathbf{p}(\mathbf{r},t)$ is assumed to depend on the 2D momentum~$\mathbf{p}$ and the 2D position~$\mathbf{r}$, while the dynamics in the third dimension is assumed to be completely frozen by a tight confinement. Then the kinetic equation reads
\begin{equation}
\frac{\partial{f}_\mathbf{p}}{\partial{t}}
+ \mathbf{v}_\mathbf{p}\cdot\frac{\partial{f}_\mathbf{p}}{\partial\mathbf{r}}
+ \mathbf{F}\cdot\frac{\partial{f}_\mathbf{p}}{\partial\mathbf{p}}
= \mathrm{St}[f],
\label{eqn:kinetic}
\end{equation}
where $\mathbf{F}$ is an externally applied force, and $\mathbf{v}_\mathbf{p}=\partial\epsilon_\mathbf{p}/\partial\mathbf{p}$ is the electron group velocity determined by the energy dispersion $\epsilon_\mathbf{p}$. The collision integral on the right-hand side, written in the Born approximation,
\begin{equation}\label{eqn:Stoss}
\mathrm{St}[f]=
2\pi{n}_\mathrm{i}\int\frac{d^2\mathbf{p}'}{(2\pi)^2}\,
|U(\mathbf{p}-\mathbf{p}')|^2\,
\delta(\epsilon_\mathbf{p}-\epsilon_{\mathbf{p}'})
\left(f_{\mathbf{p}'}-f_\mathbf{p}\right),
\end{equation}
is determined by the impurity concentration $n_\mathrm{i}$ and the Fourier transform of the impurity potential $U(\mathbf{p}-\mathbf{p}')$. Beyond the Born approximation, $|U(\mathbf{p}-\mathbf{p}')|^2$ should be replaced by the exact scattering amplitude, properly normalized.

In the absence of perturbations, the electrons are assumed to have the Fermi-Dirac distribution determined by the Fermi energy~$\epsilon_F$ and the temperature~$T$:
\begin{equation}
f^\mathrm{eq}_\mathbf{p}=\frac{1}{e^{(\epsilon_\mathbf{p}-\epsilon_F)/T}+1}.
\end{equation}
If a perturbing electrostatic potential $\varphi_{\mathbf{q},\omega}\,e^{i\mathbf{q}\mathbf{r}-i\omega{t}}+\mbox{c.c.}$ is applied to the 2D system (again, we neglect its dependence on the third coordinate), it enters Eq.~(\ref{eqn:kinetic}) via the associated electrostatic force $\mathbf{F}=-ie\mathbf{q}\varphi_{\mathbf{q},\omega}\,e^{i\mathbf{q}\mathbf{r}-i\omega{t}}+\mbox{c.c.}$. To the linear order in the perturbation, the distribution function can be sought in the form
\begin{equation}
f_\mathbf{p}(\mathbf{r},t)=f^\mathrm{eq}_\mathbf{p}+
e\varphi_{\mathbf{q},\omega}\,g_\mathbf{p}\,
e^{i\mathbf{q}\mathbf{r}-i\omega{t}}+\mbox{c.c.},
\end{equation}
where $g_\mathbf{p}$ is position- and time-independent and satisfies the following linear integral equation:
\begin{align}
&-i\omega{g}_\mathbf{p}+i\mathbf{q}\mathbf{v}_\mathbf{p}g_\mathbf{p}
-i\mathbf{q}\mathbf{v}_\mathbf{p}\,
\frac{\partial{f}^\mathrm{eq}}{\partial\epsilon_\mathbf{p}}{}\nonumber\\
{}&{}=2\pi{n}_\mathrm{i}\int\frac{d^2\mathbf{p}'}{(2\pi)^2}\,
|U(\mathbf{p}-\mathbf{p}')|^2\,
\delta(\epsilon_\mathbf{p}-\epsilon_{\mathbf{p}'})
\left(g_{\mathbf{p}'}-g_\mathbf{p}\right).
\label{eqn:linearBoltzmann}
\end{align}
An explicit solution of this equation can be found only in the case of short-range impurities when $U(\mathbf{p}-\mathbf{p}')$ does not depend on momentum. In this case the collision integral reduces to the relaxation time approximation:
\begin{equation}\label{eqn:tauapprox}
\mathrm{St}[f]=\frac{\overline{f_\mathbf{p}}-f_\mathbf{p}}\tau,
\end{equation}
where the overbar denotes the average over the momentum directions on a constant energy surface,
\begin{equation}
\overline{f_\mathbf{p}}\equiv
\frac{\int{d}^2\mathbf{p}'\,f_{\mathbf{p}'}\,\delta(\epsilon_{\mathbf{p}'}-\epsilon_{\mathbf{p}})}{\int{d}^2\mathbf{p}'\,\delta(\epsilon_{\mathbf{p}'}-\epsilon_{\mathbf{p}})},
\end{equation}
and the relaxation time and the density of states per unit area are given by (the factor of 2 in front of the integral takes into account two spin projections)
\begin{equation}
\frac1\tau=\pi\nu{n}_\mathrm{i}|U|^2,\quad
\nu=2\int\frac{d^2\mathbf{p}'}{(2\pi)^2}\,
\delta(\epsilon_\mathbf{p}-\epsilon_{\mathbf{p}'}).
\end{equation}
Then, Eq.~(\ref{eqn:linearBoltzmann}) gives
\begin{equation}
g_\mathbf{p}=\frac{i\mathbf{q}\mathbf{v}_\mathbf{p}(\partial{f}^\mathrm{eq}/\partial\epsilon_\mathbf{p})+(1/\tau)\,\overline{g_\mathbf{p}}}{i\mathbf{q}\mathbf{v}_\mathbf{p}-i\omega+1/\tau}.
\end{equation}
Averaging both sides over the momentum directions, one obtains a closed equation for $\overline{g_\mathbf{p}}$ and readily finds
\begin{equation}
\overline{g_\mathbf{p}}=\frac{1+(i\omega\tau-1)S}{1-S}
\frac{\partial{f}^\mathrm{eq}}{\partial\epsilon_\mathbf{p}},
\end{equation}
where $S$ stands for the following angular average:
\begin{align}
S\equiv{}&{}\overline{(1-i\omega\tau+i\mathbf{q}\mathbf{v}_\mathbf{p}\tau)^{-1}}
\nonumber\\
={}&{}\int_0^{2\pi}\frac{d\phi}{2\pi}\frac{1}{1-i\omega\tau+iv_Fq\cos\phi}\nonumber\\
={}&{}\frac{1}{\sqrt{(1-i\omega\tau)^2+(v_Fq\tau)^2}}.
\end{align}
The last two lines were written under the assumption of an isotropic dispersion $\epsilon_\mathbf{p}$.
Finally, since the electron density is given by
\begin{equation}
\rho(\mathbf{r},t)=2\int\frac{d^2\mathbf{p}}{(2\pi)^2}\,f_\mathbf{p}(\mathbf{r},t)
\end{equation}
(again, the factor of 2 takes care of the spin multiplicity), the density response function can be found as
\begin{equation}
\Pi(\mathbf{q},\omega)=2\int\frac{d^2\mathbf{p}}{(2\pi)^2}\,g_\mathbf{p}.
\end{equation}
Collecting all factors and assuming that $-\partial{f}^\mathrm{eq}/\partial\epsilon_\mathbf{p}$ is a narrow peak around the Fermi energy of width $\sim{T}$, so that the energy dependence of $\nu$ and $v_F$ can be neglected, we arrive at Eq.~(\ref{eqn:polarizability}).

If $U(\mathbf{p}-\mathbf{p}')$ is momentum dependent, no closed solution can be obtained even in the simplest of isotropic scattering when the scattering amplitude depends only on the difference $\phi-\phi'$ of the polar angles $\phi,\phi'$ associated to the 2D vectors $\mathbf{p},\mathbf{p}'$. Indeed, in this case the solution can be sought as a sum over polar harmonics, $g_\mathbf{p}=\sum_mg_me^{im\phi}$, which are the eigenfunctions of the collision integral. Different harmonics do not separate because of the second term in Eq.~(\ref{eqn:linearBoltzmann}), which is responsible for the spatial dispersion of the conductivity $\sigma(\mathbf{q},\omega)$. Only when the spatial dispersion is neglected, the Drude conductivity $\sigma(\omega)$ can be written in terms of the transport relaxation time $\tau_1$, determined by the first eigenvalue of the collision integral, $-1/\tau_1$. Otherwise, the result contains all eigenvalues $-1/\tau_{m>0}$. Still, qualitatively, it is $\tau_1$ that determines the relevant time scale: in the limit of Eq.~(\ref{eqn:tauapprox}) all $\tau_{m>0}=\tau$, while in the opposite limit of small-angle scattering $\tau_m$ quickly grows with~$m$, so high harmonics are suppressed and the result is determined by the first few $\tau_m$'s.

\bibliography{HeatTransfer}

\end{document}